\documentclass[12pt]{article} 
\usepackage{amssymb}
\usepackage{graphicx}
\usepackage{url}
\begin{document}  
\begin{center} 
{\large Digitized Waveform Processing for Fast Timing}


				Sebastian White, \it{CERN/University of Virginia}

\end{center}

\begin{abstract}
	The prospect of pileup induced backgrounds at the High Luminosity LHC (HL-LHC) has
stimulated intense interest in technology for charged particle timing at high rates\cite{challenge}. 
In this paper I report on a framework for fast timing sensor and related electronics development
used primarily within the context of PICOSEC
\footnote{
 PICOSEC Collaboration:
European Organization for Nuclear Research (CERN), Geneva 1211, Switzerland;    
University of Virginia, Charlottesville, Virginia 22903, USA;
CEA, IRFU, Centre d'\'{E}tudes Nucl\'{e}aires de Saclay, Gif-sur-Yvette 91191, France ;
 Univ. of Santiago de Campostela, Santiago de Campostela, Spain;
 University of Science and Technology of China, Hefei, China;
 LIP, Lisbon, Portugal;
 Helsinki Institute of Physics, Helsinki, Finland;
 Aristotle University of Thessaloniki, Greece;
 NCSR Demokritos, Athens, Greece}.
 Our collaboration accumulated a large ($\sim$fewTbyte) set of waveforms from timing sensors based on
 MicroChannel Plates(MCP), MicroPattern Gas Detectors(MPGD) and capacitive readout Avalanche Diodes (aka HFS) 
 with typically 20-40 GSa/s waveform sampling. We have reported charged particle time resolutions of 3.6, 24 and 20 picoseconds, respectively for these sensors.
In this paper I discuss some of the tools developed during this activity for the processing of waveforms
digitized at sampling rates ranging from 40 MHz (ATLAS ZDC) to 40 GHz (PICOSEC).
	\end{abstract}

\section{Introduction}
	As high bandwidth digital oscilloscopes with reasonable data acquisition capability and inexpensive waveform digitizers have come on the market (eg. commercial variants of the 
DRS4 and SAMPIC, developed at PSI and Orsay/Saclay, respectively), digital waveform data from new devices for sub 100 picosecond timing has become increasingly common.

	One precendent for this in High Energy Physics has been the realization in the '90's that the time structure of Calorimeter waveforms contained useful information about the shower
electromagnetic fraction\cite{Caldwell,SPACAL}. Perhaps more relevant is the introduction of ``Optimal Filtering" by Cleland et al.\cite{Cleland}, which is at the heart of the ATLAS LAr calorimeter 
readout. Not only does the best processing of 5 or so samples give the optimal performance for timing and energy measurement, but the optimum can adapt to changing conditions
(ie higher pileup as the LHC matures).

	At this early stage in the development of fast sensor systems one might hope that a similar analysis of fast timing waveforms could be used to guide the sensor technology as well as leading to
the best strategies for the electronics design. At the present time it seems likely that R$\&$D carried out using digitized waveforms will lead to the design of less data intensive readout but there
are a few enthusiasts who would hold out for fully digitized waveforms at $\ge1$GSa/s.

	In this paper I report on various examples of waveform analysis techniques developed over the past 10 years, beginning with the ATLAS ZDC project which, somewhat surprisingly, achieved a time resolution
on hadronic showers of $\le 100$ picoseconds\cite{Otranto,mymathematica}. I then discuss the PICOSEC project, where similar techniques led to the optimization and succesful simulation of MicroMegas based charged particle
timing detector with $\sigma_t \le 24$ picoseconds. Lastly I describe a capacitive readout avalanche diode (aka HyperFast Silicon-HFS) where work continues today to combine waveform analysis with commercial tools (eg SILVACO) to fully model the contributions to time resolution($\sigma_t \le 20$ picoseconds), including Landau/Vavilov fluctuations. Some of this work was carried out collaboratively with Wolfram Research and I point out related tools developed using Mathematica\textsuperscript{TM}.
		
\begin{figure}
\centering
\centerline{\includegraphics[width=7cm, height=5cm]{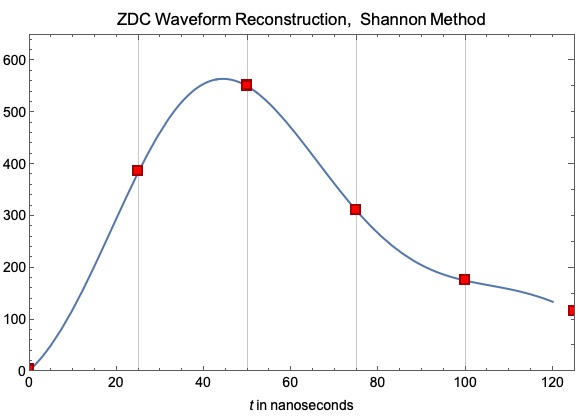}}
\caption{Typical ZDC calorimeter pulse sampled at 40 MHz (red) and reconstructed waveform. A small ad hoc correction is derived from scan data, where the signal is scanned in 1 nanosecond steps relative to the LHC clock. }
\label{fig:shannon}       
\end{figure}

\section{Calorimeter Waveforms-ATLAS ZDC}
	Although the ZDC trigger signals are brought to the ATLAS control room on short, fast cables, the signals used for digitization, due to cost and infrastructure considerations, use 320m long ethernet cables  which give poorer timing properties. As a result the signal has a frequency spectrum which rolls over at about 100 MHz ($\omega_{MAX}$).

	In his classic paper deriving the sampling theorem\cite{Shannon}, Claude Shannon uses the interpolation formula:
		
\begin{equation}
f(t)=\sum_{n=-\infty}^{\infty}x_n\frac{Sin(\pi(t/T-n))}{\pi (t/T-n)}
\end{equation}
where $x_n$ are sampled values of the waveform at time, t/T=n, and shows that once the sampling interval, T, is smaller than 1/(2$\times\omega_{MAX})$ this formula gives perfect reconstruction of the waveform. In the case of the ZDC there are a total of 7 sampling points spaced 25 nsec apart so we are not in the limit of perfect sampling. Nevertheless, based on our experience with picosecond timing of fast signals, using digital scopes, which implement eqn. 1 on-chip, we found that this formula gave the best possible timing resolution for sparse sampling. Therefore, we decided to use this elegant interpolation formula to reconstruct the time and energy of ZDC waveforms\cite{Otranto}. In ref.\cite{mymathematica} the 1 nsec delay scan mapping the non-linearities in  response algorithm, when using our digitization in the ATLAS L1calo pre-processor electronics may be found. 

\section{Filtering}
In the calorimetry example above, good signal timing with only a few samples is possible because noise considerations don't apply. High energy showers in the ZDC produce large light signals in the Quartz/Tungsten calorimeter which is detected by conventional PMTs. In general, however, finer signal sampling can provide a tool for noise reduction. Since the limiting time jitter (ie from Constant Fraction Timing) is given by:
\begin{equation}
\delta_T=t_{Rise}/SNR
\end{equation}
the optimal signal processing would remove high frequency noise, which degrades the signal leading edge, without increasing $t_{Rise}$. 
\begin{figure}
\centering
\centerline{\includegraphics[width=0.7\textwidth, height=5cm]{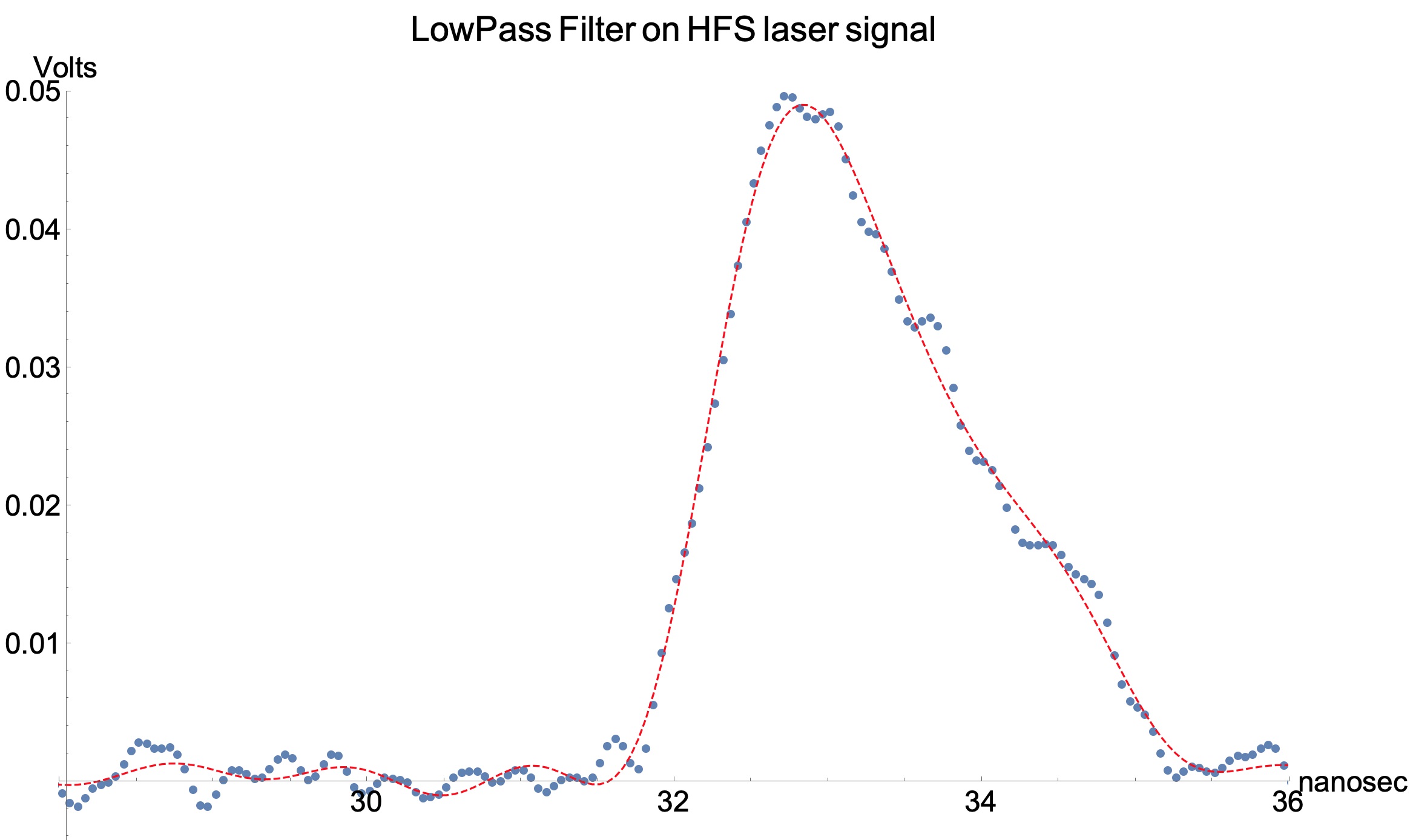}}
\caption{HFS output pulse before and after (dashed red) applying a lowpass ``brick-wall" filter.}
\label{fig:Bandpass}       
\end{figure}

	This can be easily carried out by applying successively, discrete Fourier Transform functions and then, after masking undesired frequencies, the Inverse Discrete Fourier Transform -aka ``brick-wall filter" as illustrated in  figure\,\ref{fig:Bandpass}.

\begin{figure}
\centering
\centerline{\includegraphics[width=\textwidth, height=5cm]{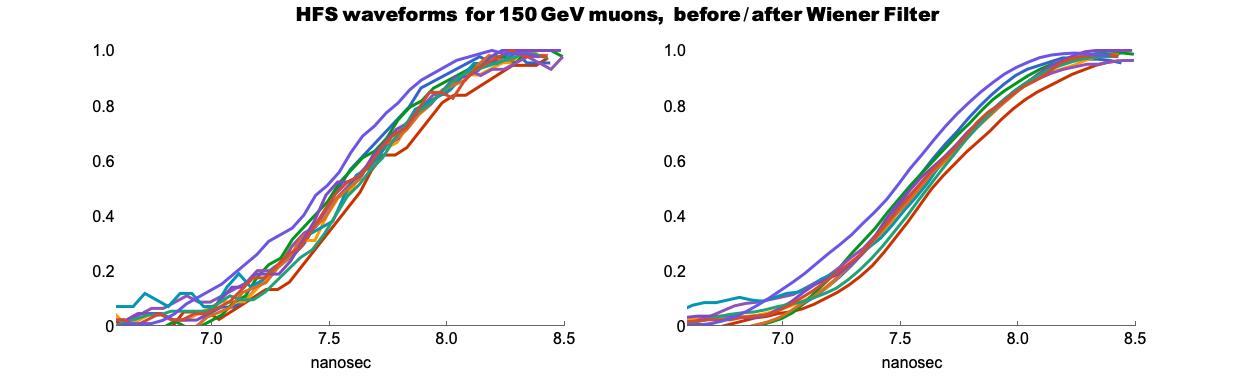}}
\caption{Application of Wiener Filter with target noise power level matching data for a sample of HFS (operated at modest internal gain) signals in a high energy muon beam. Waveforms are normalized to peak amplitude sample.}
\label{fig:Wiener}       
\end{figure}

	Several alternatives to bandpass filtering exist which can, for example, incorporate knowledge of the relative additive noise power. An example of a Wiener filter is shown in the right hand panel of figure\,\ref{fig:Wiener}.
	
\begin{figure}[!htp]
\includegraphics[width=.49\textwidth]{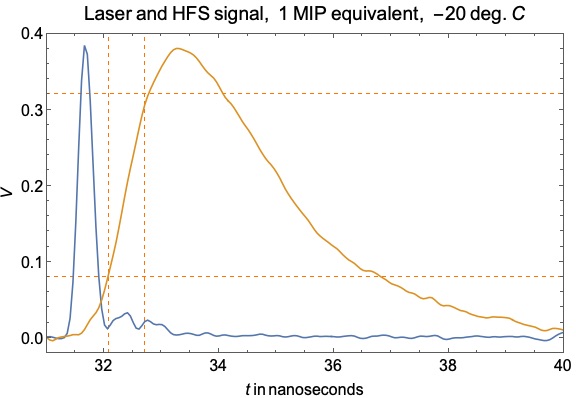}
\includegraphics[width=.49\textwidth]{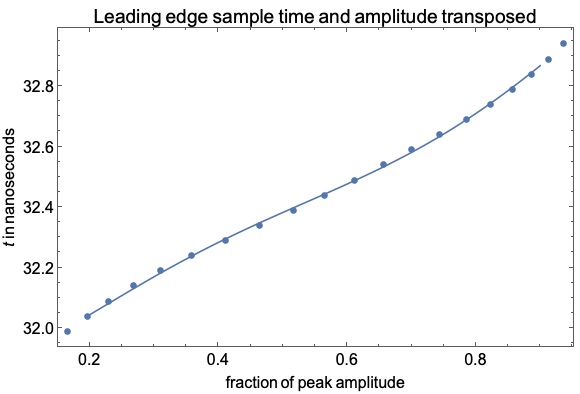}
\caption{HFS Silicon response to a fast laser pulse(left) and the normalized and transposed waveform leading edge with superposed polynomial fit (right).}
  \label{fig:HFlaser}
\end{figure}

		In many problems where local fits of the waveform are applied to extract the signal time the fit itself finesses the contribution to jitter from high frequency noise. This is illustrated in figure\,\ref{fig:HFlaser} which shows the HFS Silicon response to a fast laser pulse delivering the same signal amplitude as a minimum ionizing particle. 
	
	In order to derive the signal arrival time the waveform is renormalized to the peak amplitude (in order to derive a Constant Fraction time- eg $20\%$) and then time and amplitude are transposed. A convenient way to do this is
to perform a power law fit to the leading edge. Clearly in this case the fitting procedure reduces the contribution of high frequency noise.
	
	The limited adc resolution of sampling scopes (eg commonly 8-bit) can contribute an effective ``digital noise". One way to mitigate this, where higher sampling rates are available, is to over-sample (ie beyond the Nyquist rate) and then average neighboring samples. 
	
	In some cases incremental improvements in signal timing, especially when noise is an issue, can arise from combining eg. 20, 30 and 40$\%$ constant fraction times\cite{breton}. This is an easy exercise to perform with digital waveforms.
\section{Guiding Sensor Development}

	Waveform analysis can also provide a useful tool in sensor and electronics development as we discuss below. 
	
	The PICOSEC MicroMegas detector\cite{NIMpico} consists of an entrance window (acting as a Cerenkov radiator for charged
particle detection) coated with a UV sensitive photocathode, which also acts as the negative electrode of a two stage MicroMegas Gas detector. The first (Drift) stage is only 0.2 mm thick and operated at high enough field that drift photoelectrons 
generate significant impact ionization (and hence gain). A charged particle traversing the detector produces roughly 10 photoelectrons which undergo gas amplification in the two stages. At large gain, the dominant contributor to timing jitter is
(longitudinal) diffusion of drifting photoelectrons in the gas. 

	Because of the high drift field used, the mean free path to first Townsend multiplication is shorter than the drift gap. Effectively, the single photoelectron time diffusion before this ionization step limits the time jitter in the final output
signal. Furthermore it was found that signal arrival time varied with the distance to first ionizing collision in the gas ( and hence the gain of the drift stage).

\begin{figure}[!htp]
\includegraphics[width=.49\textwidth]{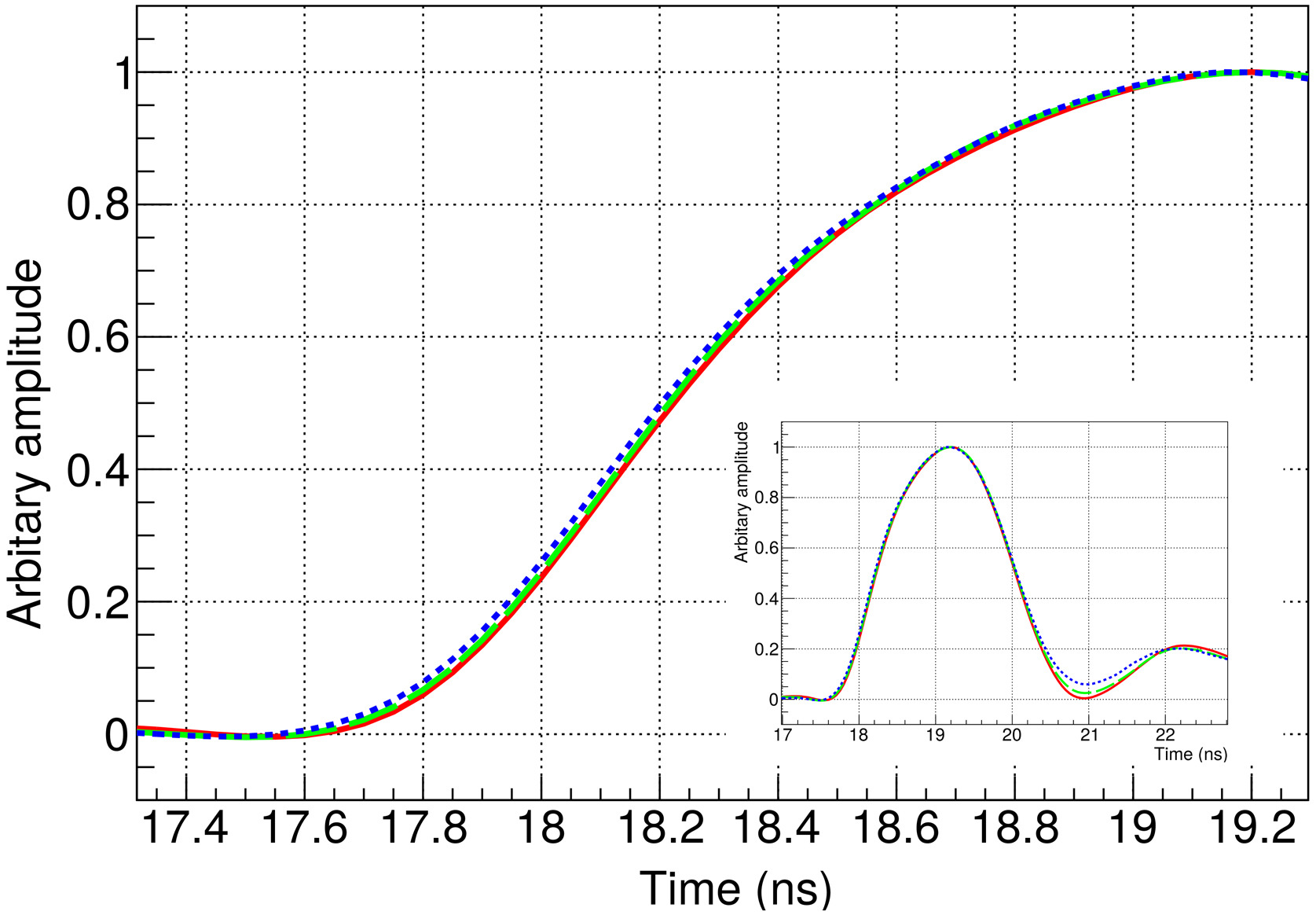}
\includegraphics[width=.49\textwidth]{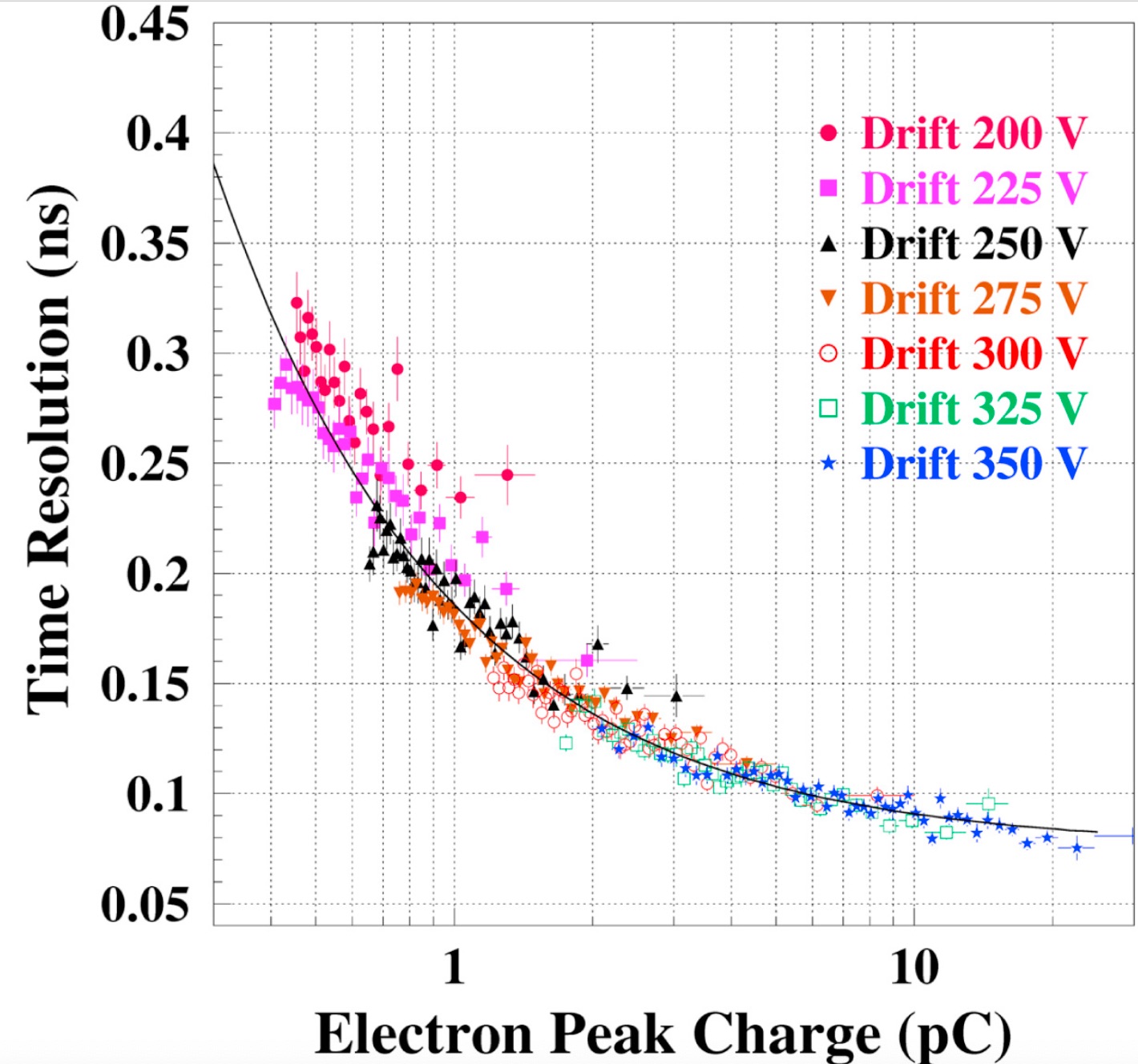}
\caption{PICOSEC single photoelectron pulses of different overall amplitude renormalized to show similarity in shapes (left). Arrival time jitter of single photoelectron pulses displayed vs. Signal amplitude(right). The same fluctuations affecting the gain in the drift region
also determine timing characteristics. Not shown here is the time resolution ($\sigma_t=  24$ picocec) observed for 150 GeV muons\cite{NIMpico} .}
 \label{fig:LIDYL}
\end{figure}

	Aspects of this detector physics underlying the time jitter of PICOSEC were elucidated by a program studying single photoelectron response at the (IRAMIS/SLIC, CEA) laser facility. Waveform analysis played a critical role as illustrated in figure\,\ref{fig:LIDYL}
The left panel shows that the signal shape is invariant with amplitude and the right panel shows that the overall jitter in arrival time of the single photoelectron signal depends solely on the overall charge. This is traceable to the distance travelled to first 
Townsend multiplication in a given event. 

	The more surprising feature of these data is that the signal \underline{arrival time} also varies with the distance to first collision. It is not trivial that the effective drift velocity is different for the initial photoelectron and the cloud of multiplied electrons. Were it not
	for the waveform data we would have suspected an artifact due to changing pulse shape.

\section{Tools for Collaboration on Front End Design}

	Inevitably the PICOSEC sensor development benefitted from close collaboration with various members of the CERN microelectronics group\footnote{particular thanks are due to Jan Kaplon and Philippe Farthouat and James Rouet.}, Eric Griesmayer (CIVIDEC) and, for most of the project, Mitch Newcomer, with whom we worked most closely on the development of a SiGe front end ASIC. 
	
\begin{figure}
\centering
\centerline{\includegraphics[width=\textwidth, height=8cm]{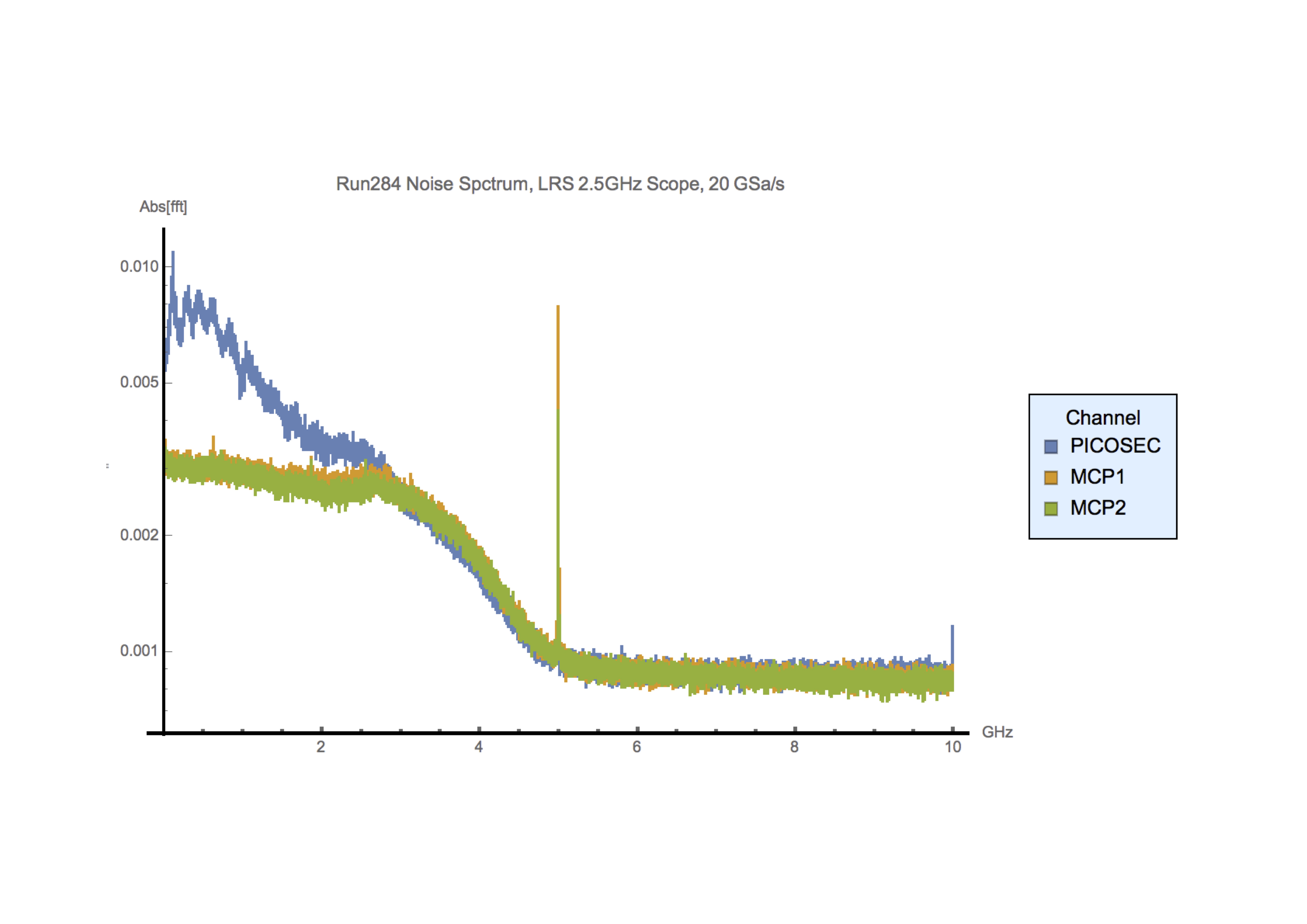}}
\caption{Fourier analysis of noise spectrum from fast timing detectors in the SPS test beam.}
\label{fig:noisefft}       
\end{figure}

	In this collaboration digital signal processing/analysis has been key in validating the performance with test devices. In one example, shown in figure\,\ref{fig:noisefft} we extract the noise power spectrum from data taking in the H4 test beam at the CERN SPS. This spectrum shows the added noise from the CIVIDEC ``C2" amplifier on the PICOSEC micromegas fast timing detector relative to the MCP channels, where the noise is dominated by the oscilloscope input noise. 
	
	Erich Giesmayer (CIVIDEC) reviewed these results and concluded ``The C2 has a 3 dB bandwidth of 1.4 GHz, this you can see from the spectrum.
We made a simulation of the noise voltage of the C2 with Spice, based on noise-measurements.
This simulation correlates well with your measurements.  The dominant term is the transfer function of the amplifier. The peak at 5 GHz is an artifact due to the anti-aliasing filter of the scope".
	In many cases a good engineer will be far more interested in hearing these details than the bottom line, which is the business of the managers.
	
\section{Interoperability}

	The material presented above was obviously developed within a particular software framework. Mathematica\textsuperscript{TM} happens to be the one that I have chosen but other options are, obviously, available. Nevertheless it is worth noting the specific 
projects that have suited themselves particularly to the style of sensor development in PICOSEC. Some of these are reported in ref.\cite{diana}. To give one example: for certain detectors- notably the MicroChannelPlate PMT, using the entrance window as a Cerenkov radiator- an appropriate tool for timing is signal modeling. In ref.\cite{diana} we describe a Cloud app we deployed which, when presented with Lecroy ``zipped" scope binary files analyzed the waveforms and reported back $\sim 7$ picosecond time jitter between 2 channels (only slightly larger than the customized constant fraction analysis)- not a bad result!

\section{Prospects}

	The Fast timing upgrades of both CMS and ATLAS are based on Silicon sensors with internal gain (Avalanche Diodes operated in Geiger mode-aka SiPM photodetectors- and Low Gain avalanche diodes). Although the former have undergone a rapid developement in the past years- partly driven by applications outside high energy physics, silicon sensors which directly measure time of arrival of charged particles (by their ionization signal) have been evolving relatively slowly over the past 20 years (partly due, no doubt, to the paucity of applications outside of high energy physics). This evolution has lately focused on the severe radiation environment of the HL-LHC applications but will hopefully be generalized. Unlike the PICOSEC example given above, further work is needed to integrate the commercial simulation tools for Silicon with internal gain (where the choice of impact ionization models is often simply a menu). Some of this research continues within a subactivity of the CERN Silicon Sensor Development lab.
	
	A more direct connection with the HL-LHC upgrades concerns the electronics strategy for the CMS Barrel Timing Layer. This sensor technology ($\sim3$mm thick LYSO crystals readout with SiPM photosensors) will be digitized by a  $\sim 1/2$ million channel ``TOFHIR' electronics- large, even by the standards of PET scanners. Unlike PET, where the total energy deposit is an important tool for preserving image quality, the CMS timing layer uses the energy to provide `walk correction" for a low threshold timing discriminator- aka NINO. So a logical question for the TOFHIR design is ``would information from timing discriminators at 2 different thresholds provide as effective a walk correction?" The preliminary answer, from waveform analysis, is ``yes"\cite{sasha}.
\subsection{Acknowledgement}
	This work received partial support through the US CMS program under DOE contract No. DE-AC02-07CH11359.

 
\end{document}